%% file: adhs-2024-ddstl2.tex
\documentclass{ifacconf}

\usepackage{graphicx}      % include this line if your document contains figures
\usepackage{natbib}        % required for bibliography
\usepackage{tikz}
\usetikzlibrary{automata,positioning,arrows}
\usepackage[utf8]{inputenc}

\usepackage{amsfonts}
\usepackage{amsbsy}
\usepackage{mathtools}
\usepackage{arydshln}
\usepackage{algorithm,algorithmicx,algpseudocode}
\usepackage[mathscr]{euscript}
\usepackage{verbatim}

\newtheorem{definition}{Definition}

\newtheorem{proposition}[thm]{Proposition}
\newtheorem{remark}{Remark}
\usepackage{enumitem}

\renewcommand{\cite}[1]{\citep{#1}}
\usepackage[framemethod=tikz]{mdframed}

\newmdtheoremenv[
hidealllines=false,
leftline=true,
innerleftmargin=1pt,
innerrightmargin=1pt,
innertopmargin=2pt,
]{problem}{Problem}

\newcommand{\showlongversion}{1}

\DeclareMathOperator*{\argmin}{arg\,min}
\input{Commands}

\newcommand{\Birgit}[1]{{\color{black} #1}}
\usepackage{subfig}

\newcommand{\new}[1]{\color{black}  #1\color{black}}
\begin{document}
	\begin{frontmatter}
\title{Direct data-driven control with signal temporal logic specifications\thanksref{footnoteinfo}}

\thanks[footnoteinfo]{This work is supported by
	the Dutch NWO Veni project CODEC (18244) and the Horizon Europe EIC project SymAware (101070802).}

\author[First,Third]{B.C.~van Huijgevoort} 
\author[First]{C. Verhoek} 
\author[First,Second]{R. T{\'o}th}
\author[First]{S. Haesaert}

\address[First]{Eindhoven University of Technology, the Netherlands}
\address[Third]{Max Planck Institute for Software Systems, Germany}
\address[Second]{The Institute for Computer Science and Control, Hungary} 
	
	\begin{abstract}
	Most control synthesis methods under temporal logic properties require a model of the system, however, identifying such a model can be a challenging task. In this work, we develop a direct data-driven control synthesis method for temporal logic specifications, which does not require this explicit modeling step, capable of providing certificates for the general class of linear systems. After collecting a single sequence of input-output data from the system, we construct a data-driven characterization of the behavior. Using this characterization, we synthesize a controller, such that the controlled system satisfies a (possibly unbounded) temporal logic specification. The underlying optimization problem is solved by mixed-integer linear programming. We demonstrate the applicability of the results through simulation examples.
	\end{abstract}

\end{frontmatter}

\section{Introduction}
\Birgit{To achieve reliability of safety-critical systems, such as autonomous vehicles and power grids, it is crucial to obtain \emph{formal guarantees} on their behavior. The use of \emph{temporal logics} in control synthesis enables the development of controllers that give such formal guarantees \cite{tabuada2009verification,belta2017formal}, while also allowing the description of complex behavior required for such systems \cite{baier2008principles}. The automatic construction of a provably correct controller based on these specifications and the model of the system is referred to as \textit{correct-by-design} control synthesis. 
 Most of these methods \cite{tabuada2009verification,belta2017formal} 
explicitly assume the availability of a completely known analytic model of the system. 
However, in many realistic physical applications, the model of the considered systems is not known or not fully known.}

Due to the increasing complexity of systems, obtaining an accurate model has become a challenging task in practice \cite{hjalmarsson2005experiment}. Additionally, some part of the system is always left unmodeled due to limitations of first-principles-based modeling or uncertainty of the estimated model. These modeling errors propagate into the control design and result in unwanted behavior. A way to circumvent this issue is to directly synthesize a controller from data. By doing so, we avoid the possibility that an approximation error occurs in the modeling process. The interest in such \emph{direct data-driven} control synthesis techniques is increased by the huge success of e.g. reinforcement learning \cite{sutton1999reinforcement}. Unfortunately, those approaches usually require a large amount of data \cite{recht2019tour}, while the approach developed in this paper requires only a single sequence of input-output data.
More specifically, in this work, we bring down correct-by-design control synthesis to the level of data by using a data-driven method 
to directly synthesize a controller \Birgit{for both bounded and unbounded specifications}. 

{\noindent\bfseries{Literature.}}
A common methodology in control design is to apply first-principles-based modeling combined with parameter estimation to obtain a model and use that model to design a (model-based) controller. Similar approaches are used in correct-by-design control synthesis in 
\cite{haesaert2017data,filippidis2016control}, where 
after obtaining a description of the continuous-state model, one resorts to abstraction-based or abstraction-free methods to construct a provably correct controller \cite{belta2017formal,tabuada2009verification}.
Unfortunately, \new{for physical, causal systems this modeling step is typically approximate and these approximation errors get even worse } for complex systems. \new{Furthermore, there is no separation principle between these steps. More precisely, the parameters that best fit the data may not yield the best model for control design \cite{dorfler2023data}. }
Instead of starting with a first-principle-based continuous-state model, there also exist correct-by-design control synthesis methods that use data to obtain a finite-state abstract model \cite{ devonport2021symbolic, kazemi2022data, lavaei2022constructing} and approaches that use reinforcement learning to learn a control policy such that a specification is satisfied  \cite{kapoor2020model, kalagarla2021model, kazemi2020formal}. 
\Birgit{In all of these approaches, the measurement data consists of \emph{state} trajectories, while the method presented in this paper refrains from using information about the state of the system and requires only input-output data. More specifically, 
we show conditions under which  this modeling step can be omitted and a more simple straightforward direct data-driven approach is possible.}

Recently, direct data-driven control has gained a lot of interest, mainly because it shows promising results for analysis, simulation, and control of complex systems. Such methods are generally based on the Fundamental Lemma \cite{willems2005note} of the behavioral framework for LTI systems \cite{willems1997introduction} to obtain a full characterization of the system behavior using only measurement data. Based on this Lemma, a wide range of powerful tools have been developed in the LTI case for data-driven simulation \cite{markovsky2008data}, performance analysis \cite{koch2021provably,van2022data}, and control with closed-loop performance guarantees \cite{markovsky2007linear,coulson2019data,berberich2020data}.

\noindent{\textbf{Contributions. }
To the best of our knowledge, direct data-driven control has never been used to automatically construct a controller that guarantees the satisfaction of a temporal logic specification, even though this allows the design of a \emph{provably correct} controller and can be applied to satisfy \emph{complex specifications} that describe rich behavior of a system \new{that can generally not be expressed though e.g. state inequalities}.
In this work, we contribute to the existing literature as follows:
\begin{itemize}[noitemsep,topsep=0pt]
	\item For bounded signal temporal logic (STL) specifications, we develop a direct data-driven approach to automatically synthesize a controller using only input-output data. % and prove its soundness and completeness
	\item We provide an algorithm that performs the control synthesis.
	\item We extend the direct data-driven approach to \emph{unbounded} STL specifications by introducing a novel loop constraint for input-output signals and give the corresponding algorithm.
\end{itemize}

\new{This paper combines the research areas of temporal logic and direct data-driven control and is restricted to linear systems and noise-free data. However, the ongoing trends in direct data-driven control include developments towards nonlinear systems \cite{verhoek2021data,verhoek2021fundamental} and noisy data \cite{breschi2023data}, and allows extensions of this paper to this broader range of systems. }
This paper is structured as follows. We start by introducing the relevant theory and defining the considered problem setting. In Section~\ref{sec:dataDriven}, we introduce a data-based representation of the behavior of LTI systems, which we use for control design subject to a \Birgit{(possibly unbounded)} temporal logic specifications in Section~\ref{sec:TLcontrol}. 
In Section~\ref{sec:resultss}, we employ our method in multiple case studies, and we 
end with conclusions on the proposed methodology in Section~\ref{sec:discussion}.  \ifthenelse{\equal{\showlongversion}{1}}{}{The proofs of this paper are given in the online version \cite{}.}

\section{Problem statement}
In this section, we introduce the relevant theory followed by a formal problem statement 

\subsection{Notation}
We define the set of natural numbers by $\N$, the set of integers by $\Z$, and	the set of real numbers by $\mathbb{R}$.
In the sequel, we denote with finite intervals $[a,b]$ the ordered sequence $\{a, a+1, \ldots, b\}$ with $a<b$ and $a,b\in \N$.
Similarly we use $\mathbf{z}$ to denote the infinite sequence $\{\ldots, z_{-2}, z_{-1},z_0,z_{1}, z_{2},\ldots\}$, while  $\mathbf{z}_t=\{z_t,z_{t+1}, z_{t+2},\ldots\}$.  	We use  $\mathbf{z}_{[0,N]}$ to denote the finite sequence $\{z_0,z_1, \ldots, z_N\}$ or its stacked version\footnote{Sequences can be stacked either row-wise or column-wise depending on the context.}.

\subsection{Discrete-time dynamical systems}

As  in Def.~1.3.4 in \cite{willems1997introduction},  we define discrete-time dynamical systems based on their behavior as follows.
\begin{definition}[Dynamical system]\label{def:dynamical}
A dynamical system $\Sys$ is defined as a triple
$	\Sys=(\T,\W,\Beh)$
with $\T$ a subset of  $\Z$, called the time axis, $\W$ a set called the signal space, and $\Beh$ a subset of $\W^\T$ called the behavior. Here, $\W^\T$ is the notation for the collection of all maps from $\T$ to $\W$.  
\end{definition}
The behavior $\Beh$ of a system is a set of trajectories or time-dependent functions  that are compatible with the system.
In this work, we define a system, its initialization, and its control design via its behavioral set. This way no exact knowledge on the structure of the system such as the dimension of the state space of the system is required.

\medskip

\noindent{\textbf{Linear Time Invariant Systems. }
In the sequel, we will focus on Linear Time Invariant (LTI) systems as defined in Defs.~1.4.1 and 1.4.2 in \cite{willems1997introduction}. 
Dynamical systems are \emph{linear} if the superposition principle holds, i.e., 
\Birgit{if trajectories $w_1,w_2 \in \Beh$, then $\forall \alpha, \beta \in \mb{R}$: $\alpha w_1 + \beta w_2 \in\Beh$.
\emph{Time-invariance} of a system implies that any time-shifted version of a trajectory in $\Beh$ is again an element of $\Beh$, i.e., $\q^\tau\Beh\subseteq\Beh$  holds for all $\tau\in\mb{T}$, where $\q^\tau w_t=w_{t+\tau}$. } 
A state-space representation is a realization of $\Beh$. Consider the generic state-space representation  of an LTI system given as 
\begin{equation}\label{eq:model}
\begin{cases}
x_{t+1} &= A x_t+B u_t \\
y_t  &= C x_t + D u_t
\end{cases}
\end{equation} with state $x_t\in \mathbb X = \mathbb{R}^{n_\mr{x}},$ input $u_t \in \mathbb U= \mathbb{R}^{n_\mr{u}}$, and output $y_t \in  \mathbb Y=\mathbb{R}^{n_\mr{y}}$. 
Matrices $A,B,$ $C$, and $D$ are of appropriate sizes. Let the signal $\w$ have a signal space $\mathbb W =\U\times \Y$ such that $\w$  is composed from input signal $\mathbf u$ and output signals $\mathbf y$, that is, $\w=(\mathbf u,\mathbf y)$. Then, representation \eqref{eq:model} defines a dynamical system $\Sys=(\Z,\W,\Beh)$ 
with behavior
\begin{equation}\notag
\Beh:= \{\w\in \mathbb W^\Z\mid \exists {\bf x} \in \mathbb{X}^\mathbb{Z}, \text{ s.t. } ({\bf y},{\bf x},{\bf u}) \textmd{ satisfy \eqref{eq:model} } \forall t\in\Z\}.
\end{equation} 	
Intuitively, we define the \emph{order} of the system  $\mbf{n}(\mf{B})$ as the minimal state dimension required in \eqref{eq:model} such that the input-output behavior of \eqref{eq:model} fully represents $\mathfrak{B}$. Similarly, we define the \emph{lag} of the system $\mathbf{l}(\mathfrak{B})$ as the minimal length of the signal $\w_{[0,N]}$ required to always fully reconstruct the state. Lastly, a behavior is considered \emph{controllable} if we can steer a system to a desired trajectory in the behavior, see \cite {MARKOVSKY202142}. 
We only require that we have rough upper bounds on the lag and the order of the system \cite{markovsky2008data}.

\medskip 

\noindent{\textbf{Finite behaviors and concatenations. }
We denote the behavior of an LTI system that only contains finite trajectories of length $T+1$ as $\Bfint{[0,T]}$, i.e.,
\[ \Bfint{[0,T]}:=\{ \w_{[0,T]}\mid\exists\,\mathbf v \in\Beh\text{ s.t. }w_t=v_t\text{ for }0\le t\le T \}. \]
We say that $\w_{[0,T]}$ is a trajectory of $\Sys$ if  $\w_{[0,T]}\in\Bfint{[0,T]}$.

For a given dynamical system $\Sys$, consider two  input-output trajectories $\w^a\in \Bfint{[0,T_a]}$ and  $\w^b\in \Bfint{[0,T_b]}$, the \emph{concatenated} trajectory  is defined as $ \w =\w^a \frown \w^b$ with $\w_{[0,T_a]} =\w^a$ and $\w_{[T_a+1, T_a+T_b+1]} =\w^b$.  
The concatenation $\w^a \frown \w^b$ might, or might not belong to $\Bfint{[0,T_a+T_b+1]}$. If it does, we say that $\w^b$ is a \emph{compatible} continuation of $\w^a$ in $\Bfint{[0,T_a+T_b+1]}$, or that $\w^a$ is a \emph{compatible} precedent of $\w^b$ in $\Bfint{[0,T_a+T_b+1]}$. Given $\w^a$, we are now interested in finding trajectories $\w^b$ such that 
$\w^b$ is a compatible continuation of $\w^a$ in $\Bfint{[0,T_a+T_b+1]}$, that is 
$\w^a \frown \w^b\in \Bfint{[0,T_a+T_b+1]}$. 

\medskip

\noindent{\textbf{Control design  and initialization. }
Given a system $\Sys$, the design of a controller corresponds to the design of a mapping that selects a sequence of inputs $\u$ that lead to a  trajectory $\w$ in $\Beh$. 
To this end, let $L>0$ and let $\u_{[0,L]}$ be a (given) control input signal. Let $\w^\mr{ini}\in \Bfint{[0,T_\text{ini}]}$ with $T_{\text{ini}}>0$ be given. If $\w_{[0,L]} = (\u_{[0,L]}, \y_{[0,L]})$  is a compatible continuation of $\w^\text{ini}$, then $\y_{[0,L]}$ is called the controlled output associated with the control input $\u_{[0,L]}$ and $\w^\text{ini}$.  We will be interested in the situation that $T_{\text{ini}}$ is such that for all initial trajectories $\w^\mr{ini}\in \Bfint{[0,T_\text{ini}]}$, the controlled output $\y_{[0,L]}$ associated with $\u_{[0,L]}$ and $\w^\mr{ini}$ is uniquely defined. The existence of such a $T_{\text{ini}}$ is guaranteed by the following proposition \cite{markovsky2008data,MARKOVSKY202142}.
\begin{proposition}[Initial Condition]\label{prop:init}
Let $ T_{\text{ini}}+1 \geq \mathbf{l}(\mathfrak{B})$. Then for any given $ \w^\text{ini} \in\Bfint{[0,T_\text{ini}]}$ and $ \u_{[0,L]} \in\left(\mathbb{R}^{\dnu}\right)^{L+1}$, there is a unique $ \y_{[0,L]} \in\left(\mathbb{R}^{\dny}\right)^{L+1}, $ such that $\w^{\text{ini}} \frown (\u_{[0,L]}, \y_{[0,L]}) \in \Bfint{[0,T_\text{ini}+L]}$. 
\end{proposition}
This means that  $\w^\text{ini} $ is the collection of past inputs and outputs that is sufficient to uniquely determine the compatible continuations of  $ \w^\text{ini}$ once the control input $\u_{[0,L]}$ is known. We, therefore, say that $\w^\text{ini} $ is an \emph{initialization} or initial condition.

\subsection{Signal temporal logic specifications and cost function} 
Consider the language of \emph{signal temporal logic} (STL) \cite{maler2004monitoring,deshmukh2017robust}
whose syntax is recursively defined as
\begin{equation} \notag
\varphi ::= \sigma \mid \neg \varphi \mid \varphi_1 \wedge \varphi_2 \mid \varphi_1 \lor \varphi_2 \mid \always_{[a,b]} \varphi  \mid \varphi_1 \until_{[a,b]} \varphi_2,
\end{equation}
starting from predicate $\sigma$ : $\sigma(y)>0$ for predicate function $\sigma:\Y\rightarrow \mathbb R$. 
\noindent The semantics are  as follows  
	$		\y_t \models \sigma $ iff $ \sigma(y_t)>0 $;
	$	\y_t \models \neg\varphi  $ iff $ \neg (	\y_t \models \varphi ) $;
	$	\y_t \models \varphi_1 \wedge \varphi_2 $ iff $	\y_t \models \varphi_1 \wedge \y_t \models \varphi_2 $;
	$	\y_t \models \varphi_1 \lor \varphi_2 $ iff $	\y_t \models \varphi_1 \lor \y_t \models \varphi_2 $;
	$		\y_t \models \always_{[a,b]} \varphi $ iff $\forall t'\in [t+a,t+b]\ \y_{t'} \models \varphi $;
	$					\y_t \models \varphi_1 \until_{[a,b]} \varphi_2 $ iff $ \exists t'\in [t+a,t+b]\ s.t. \y_{t'} \models \varphi_2   \wedge \forall t''\in [t,t'], \y_{t''} \models \varphi_1. $  
Additionally, we define the \emph{eventually}-operator as $\eventually_{[a,b]} \varphi := true \until_{[a,b]} \varphi$, which is true if $\varphi$ holds at some time on the interval between $a$ and $b$.

A (possibly infinite) trajectory $\y_t=\{y_t,y_{t+1},y_{t+2},\dots\}$ satisfies $\varphi$ denoted by $\y \models \varphi$ iff $\y_0 \models \varphi$ holds.
Furthermore, for a given input sequence $\u_0=\{u_0,u_1, u_2, \ldots\}$, and  initialization $\w^\text{ini}$, we say that  $\Sys$  satisfies specification $\varphi$, if the unique continuation $\y_0$ is such that $\y_0 \models \varphi$. 

To optimize the performance of the controller, we introduce a cost function as a performance measure. Since we use only input-output data of the system, the cost function can only be a function of $u$ and $y$. More specifically, we define a cost function $J: \mathbb{U}^{\mathbb{Z}_0^+} \times \mathbb{Y}^{\mathbb{Z}_0^+} \rightarrow \mathbb{R}$ that assigns a cost to a trajectory $(\u_0,\y_0)$. In this work, we consider cost functions that are quadratic and of the form
\begin{equation} \label{eq:cost}
J(\u_0,\y_0) =  ||\u_0 ||_R^2 + || \y_0 ||_Q^2,
\end{equation} with
$||\boldsymbol{x}_{[0,L]}||_P$ denoting the weighted-norm of a trajectory defined as $\small \sqrt{\boldsymbol{x}_{[0,L]}^\top P \boldsymbol{x}_{[0,L]}} = \sum_{t=0}^{L} \sqrt{x_t^\top P x_t}$.

\subsection{Problem statement}
We focus on direct data-driven synthesis of a controller that achieves optimal performance under STL specifications. 
More precisely, we will focus on using a single sequence of data to directly design a sound data-driven controller. 	Thus, given a finite input-output sequence $\w^\text{data}$ from an unknown LTI 
system $\Sigma= (\Z, \mathbb U\times\mathbb Y, \Beh)$,
$\w^\text{data}\in\Bfint{[0,T]}$, we want to synthesize a control input $\u_0$, such that the trajectory initialized with $\w^\text{ini}$ has a minimal cost continuation in terms of \eqref{eq:cost}  for which $\y_0\models \varphi$.
Formally this can be written as. \\[.1em]

\begin{problem}[Direct data-driven STL control] \label{prob:DDDcontrol}
Given a finite data sequence $\w^\text{data}\in\Bfint{[0,T]}$, $T_{\text{ini}}+1 \geq \mathbf{l}(\mathfrak{B})$, initial condition   $ \w^\text{ini}$, STL specification $\varphi$, and cost $J$. Solve
%solve
\begin{subequations}\label{eq:Opt}
	\begin{align}
		\arg\min_{\u_0}& \ J(\u_0,\y_0) \\
		&\textmd{s.t. } \w^\text{ini}\frown (\u_0,\y_0)\in \Beh, \\\
		&\textmd{and }  \y_0 \models \varphi .
	\end{align}
\end{subequations}
\end{problem}

\textbf{Approach.} 
In this paper, we solve Problem~\ref{prob:DDDcontrol} by performing the following steps. We start with obtaining input-output data of the system collected in $\mbf{w}^\text{data}$ 
to derive a data-driven characterization of the system.
After obtaining an initial trajectory that initializes the system, 
we rewrite Problem~\ref{prob:DDDcontrol} to synthesize a control sequence $\u_0$ that is an optimal solution to \eqref{eq:Opt}.}
\Birgit{Furthermore, we extend the approach to unbounded specifications by applying an infinite sequence of inputs that becomes periodic in finite time. }
For the sake of simplicity, we compute the entire input sequence at the beginning of the interval and then apply it fully without any re-computation. \Birgit{However, we can also formulate a receding horizon form.} 

\section{Data-driven characterization of the system}\label{sec:dataDriven}
In this section, we recall the data-driven characterization of the system based on  \cite{willems2005note}. 
\Birgit{Define the Hankel matrix of depth %row size 
$L+1$ associated with a sequence $\mathbf{z}_{[0,N]}$ with elements $z_t\in\mathbb R^{n_\mathrm{z}}$ for $t\in[0,N]$  as
\begin{equation}\label{eq:hankel}
\Hnkl_{L+1}(\mathbf{z}_{[0,N]})=\begin{bmatrix} z_{0} & z_{1} & \cdots & z_{N-L} \\ z_{1} & z_{2} & \cdots & z_{N-L+1} \\ \vdots & \vdots & \ddots & \vdots \\ z_{L} & z_{L+1} & \cdots & z_{N}\end{bmatrix}, 
\end{equation} with dimension $\Hnkl_{L+1}(\cdot) \in \mathbb{R}^{n_{\mathrm{z}} (L+1) \times N-L+1}$.} 
In order to obtain a characterization of the system using input-output data, we use the Fundamental Lemma from \cite{willems2005note}.
Suppose that we are given  $\w^\text{data}\in\Bfint{[0,T]},$ of an LTI system, then based on the linearity and shift invariance of the system, we know that for any $\alpha\in \mathbb R^{T-L+1}$ it holds that
\begin{equation}\notag
\Hnkl_{L+1}(\w^\text{data})\alpha\in \Bfint{[0,L]}. 
\end{equation}
Given that data $\w^\text{data}$ is rich enough, we also have that $\forall \w_{[0,L]}\in\Bfint{[0,L]}$ there exists an $\alpha\in \mathbb R^{T-L+1}$ such that \begin{equation}\label{eq:existg}
\Hnkl_{L+1}(\w^\text{data})\alpha=\w_{[0,L]}.
\end{equation} We will formalize these statements and the required conditions in the remainder of this section. %Furthermore, we will show that these properties hold true.
\begin{definition}\label{def:perEx}
A finite sequence $\mathbf u^\text{data}\in\mathbb{U}^{T+1}$ 
is persistently exciting of order $L+1$ if and only if 
\begin{equation*}
\mr{rank}(\Hnkl_{L+1}(\mathbf{u}^\text{data})) = (L+1) n_\mr{u},
\end{equation*} with the Hankel matrix $\Hnkl_{L+1}(\cdot)$ as in \eqref{eq:hankel}. 
\end{definition}
The Fundamental Lemma \cite{willems2005note} gives the conditions for equality \eqref{eq:existg} and shows that 
once the input is persistently exciting of order $L+1+\dnx$, the Hankel matrix of depth $L+1$ constructed from the data spans the full behavior of the LTI system, restricted to length $L+1$ trajectories. Translating this to the classical control setting, we obtain the result \mbox{from \cite{berberich2020trajectory}.}
\begin{proposition}[Fundamental Lemma for control] \label{prop:data}
Suppose $\w^\text{data} =(\u^\text{data},\y^\text{data})$
is a trajectory of an LTI system $\Sigma$, where $\u^\text{data}$ is persistently exciting of order $L+1+\dnx$. Then, $\w_{[0,L]} =  ( \u_{[0,L]},\y_{[0,L]})$
is a trajectory of $\Sigma$ if and only if there exists an $\alpha\in\mathbb{R}^{T-L+1}$ such that
\begin{equation}\label{eq:fundlemm}
\begin{bmatrix}\Hnkl_{L+1}(\u^\text{data}) \\ \Hnkl_{L+1}(\y^\text{data}) \end{bmatrix} \alpha = \begin{bmatrix}  \u_{[0,L]}  \\ \y_{[0,L]} \end{bmatrix}.
\end{equation} 
\end{proposition}

The main interpretation is that any length $L+1$ trajectory is spanned by the time shifts of the measured trajectories in the data $\mbf{w}^{\text{data}}$. Hence, the left-hand side of~\eqref{eq:fundlemm} can be interpreted as a fully data-based representation of the LTI system. This implies that when we have sufficiently informative data, we can characterize the full system behavior consisting of all length $L+1$ system trajectories.

\section{Direct data-driven temporal logic control synthesis}\label{sec:TLcontrol}
Now that we have obtained  
 a characterization of the system as in \eqref{eq:fundlemm} based on data, we can rewrite the constraints of optimization problem \eqref{eq:Opt} into a mixed-integer linear program. \Birgit{To this end, we distinguish between bounded and unbounded specifications.}

%We start by computing the required trajectory length $L+1$.
For a finite trajectory $\y_{[0,L]}$ to satisfy STL specification $\varphi$, that is $\y_0 \models \varphi$, it has to be sufficiently long. 
The necessary horizon $L+1$ can be computed based on the structure of the formula $\varphi$, cf. \cite{maler2004monitoring}. We denote the lower bound as $\|\varphi\|$ and obtain  $L+1 > \|\varphi\|.$ When $\smash{\|\varphi\| = \infty}$, we refer to the STL specification as unbounded.

\subsection{Direct data-driven  control for bounded specifications}
In optimization problem \eqref{eq:Opt} we have two constraints. The first constraint on  \emph{the system dynamics} implies that the trajectory belongs to the behavior of the model. This can be equivalently written using the characterization of the system obtained through data, i.e., as in \eqref{eq:fundlemm}.
The second \emph{STL constraint} in \eqref{eq:Opt} guarantees the satisfaction of the STL specification. This can equivalently be rewritten as a mixed-integer linear program (MILP) using the method described in \cite{raman2014model}. 
To this end, introduce the binary auxiliary variable $\zeta_t^\varphi$ associated with a set of MILP constraints such that $\smash{\zeta_t^\varphi=1}$ if $\varphi$ holds at time $t$. As before, satisfaction of a formula is considered at $t=0$, hence if $\zeta_0^\varphi = 1$, the formula is satisfied. Variable $\zeta_0^\varphi$ is computed recursively using the associated MILP constraints. We denote the set of MILP constraints associated with a given STL formula $\varphi$ as $\boldsymbol{\zeta}_t^\varphi=1$ with $t \in [0,1, \dots, L]$. 

To design a controller, we need an initial trajectory that satisfies Prop.~\ref{prop:init}. This results in splitting up the trajectory into an initialization part $\mbf w^{\text{ini}}$ and a controlled part $\mbf{w}_0$, which extends the data-driven characterization as in \eqref{eq:fundlemm}. We denote the initializing trajectory of  length $T_{\text{ini}}+1$ by $\mbf w^{\text{ini}} = (\u^{\text{ini}},\y^{\text{ini}})$.
The optimization problem~\eqref{eq:Opt} can now be written as
\begin{subequations}\label{eq:optProbBounded}
\begin{align}
\argmin\limits_{\u_{[0,L]} \in \mathbb{U}^{L+1}, \boldsymbol{\zeta}_t^{\varphi_L}} \hspace{-1cm}&\hspace{1cm}J(\u_{[0,L]},\y_{[0,L]}) \\
\text{s.t.} &  	\begin{bsmallmatrix}\Hnkl_{T_\text{ini}+L+1}(\u^\text{data}) \\ \Hnkl_{T_\text{ini}+L+1}(\y^\text{data}) \end{bsmallmatrix} \alpha = \begin{bsmallmatrix} \u^\text{ini} \label{eq:constr1B}\\ \u_{[0,L]} \\ \y^\text{ini} \\ \y_{[0,L]} \end{bsmallmatrix} \vspace{2pt} \\
&\boldsymbol{\zeta}_t^{\varphi_L} = 1 \text{ as in \cite{raman2014model}}, \label{eq:constr2B} \\
&\zeta_0^{\varphi_L} = 1, \label{eq:constr3} 
\end{align}
\end{subequations} 
This optimization problem is solvable using a MILP algorithm, e.g., with \texttt{Gurobi} \cite{gurobi}.	
The complete direct data-driven  control synthesis procedure is summarized in Algorithm~\ref{alg:controlFin}.

\begin{algorithm}
\caption{Direct data-driven control synthesis 
}
\begin{algorithmic}[1]
\State \textbf{Input:} $\w^\text{data}, \w^\text{ini}, \varphi, J$,
\State Compute  
$L$, such that $L+1 > \|\varphi\|$ 
\State  Compute $\Hnkl_{T_\text{ini}+L+1}(\w^\text{data})$ for constraint \eqref{eq:constr1B}
\State Rewrite constraint $\y_{[0,L]} \models \varphi$ to MILP constraints \eqref{eq:constr2B}
\State $\u_{[0,L]} \leftarrow $ solve optimization problem \eqref{eq:optProbBounded}.
\end{algorithmic}\label{alg:controlFin}
\end{algorithm}

\begin{figure*}[htp]
\input{loopFigure.tex} \hspace{2cm}  \input{loopFigure_lag}
\caption{Loop constraint (left) on the state and (right) on the signal $w$ composed of the input and output $(u,y)$.}
\label{fig:loopfigs}
\end{figure*}
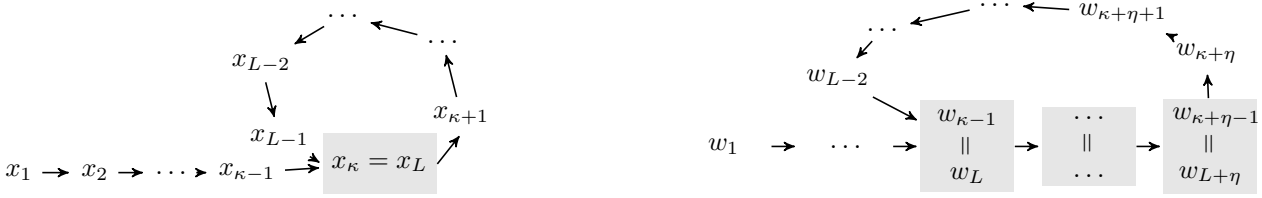

\subsection{Soundness and completeness analysis}\label{sec:SoundComplete}
In this section, we analyze the soundness and completeness of \new{optimization problem \eqref{eq:optProbBounded}. It is more common to analyze these properties for an algorithm, however, Algorithm \ref{alg:controlFin} is a numerical algorithm whose properties depend on the solver used in the last line. } %Algorithm \ref{alg:controlFin}. 
We require two assumptions. For the soundness, we make an assumption on the length of the initializing trajectory, and for the completeness we require persistence of excitation of the data as a condition.

\subsubsection{\textbf{Soundness.}}
We say that an optimization problem is sound when the following holds: \emph{If the optimization problem returns a controller, then the controlled system will satisfy the STL specification.} In the following theorem, we show that the direct data-driven control approach summarized in  \eqref{eq:optProbBounded} is sound. 

\begin{thm}[Direct data-driven control is  sound]\label{th:sound}\hspace{-0pt}
Given initialization trajectory $\mbf{w}^{\text{ini}}$ of length $ T_{\text{ini}}+1 \geq \mathbf{l}(\mathfrak{B})$,
any solution $\u_{[0,L]}$ of  optimization problem \eqref{eq:optProbBounded} 
has a corresponding output $\y_{[0,L]}$ that satisfies the STL specification.
\end{thm} 
\begin{pf} 	
 \ifthenelse{\equal{\showlongversion}{1}}{
Since $\u_{[0,L]}$ is a solution of %Algorithm~\ref{alg:controlFin}, it is a solution of 
optimization problem \eqref{eq:optProbBounded},  constraint~\eqref{eq:constr1B} is satisfied, which implies that the concatenation $\w^{\text{ini}} \frown (\u_{[0,L]}, \y_{[0,L]})$ is in the behavior of the system, i.e., $\w^{\text{ini}} \frown (\u_{[0,L]}, \y_{[0,L]}) \in \Bfint{[0,T_{\text{ini}}+L+1]}$. Following Prop.~\ref{prop:init}, $T_\text{ini}$ is large enough for $\w^\text{ini}$ to fully initialize the behavior, which implies that the output trajectory $\y_{[0,L]}$ is a unique optimal solution and, therefore, the only possible output of the system. 
Since the MILP constraints \eqref{eq:constr2B}-\eqref{eq:constr3} are also satisfied, it holds by construction that this output trajectory satisfies the STL specification $\varphi$, that is $\y_{[0,L]} \models \varphi$ holds. 
Concluding, the controller $\u_{[0,L]}$ obtained from \eqref{eq:optProbBounded} applied to the system $\mbf{\Sigma}$ will yield an output trajectory $\y_{[0,L]}$ such that $\y_{[0,L]} \models \varphi$ holds. 
}
{The proof is given in the online version \cite{}.
}
\end{pf}

\subsubsection{\textbf{Completeness.}} %
We say that an optimization problem is complete when the following holds: \emph{If a controller exists such that the controlled system satisfies the specification, the optimization problem will have a feasible solution.} To this end, we state the following.

\begin{thm}[Completeness of the optimization problem]
Given data sequence $\w^\text{data}$ from the controllable system~$\mbf{\Sigma}$ that is persistently exciting of order $\smash{T_\text{ini}+L+1+\dnx}$ according to Def.~\ref{def:perEx}. 
If there exists an input $\u_{[0,L]}$  to $\mbf{\Sigma}$, such that the corresponding output $\y_{[0,L]}$ satisfies the STL specification, then optimization problem \eqref{eq:optProbBounded} has a feasible solution. 
\end{thm}
\begin{pf}
	 \ifthenelse{\equal{\showlongversion}{1}}{
Since  $\w_{[0,L]}  = (\u_{[0,L]}, \y_{[0,L]})$ is a trajectory of $\mbf \Sigma$ and $\w^\text{data}$ is persistently exciting according to Def.~\ref{def:perEx}, we can follow Prop.~\ref{prop:data} to conclude that there exists an $\alpha$ such that \eqref{eq:constr1B} is satisfied. 
Since $\y_0$ satisfies the STL specification, the set of MILP constraints \eqref{eq:constr2B}-\eqref{eq:constr3} is satisfied. 
Since all constraints are satisfied, optimization problem \eqref{eq:optProbBounded} has a feasible solution. 
}
{The proof is given in the online version \cite{}.
}
\end{pf}

\subsection{Direct data-driven  control for unbounded specifications}
 Inspired by \cite{wolff2014optimization}, we know that an unbounded specification can be satisfied by a finite trajectory that becomes periodic in time. 
In the literature, this is applied to state trajectories, however, here we consider using input-output trajectories $\w = (\u,\y)$. 
Hence,  similar to Def. 2 in \cite{wolff2014optimization} for state trajectories, we define an \emph{$(L_F,\kappa)$-loop} for input-output trajectories as follows.
\begin{definition} \label{def:Lkloop}
A trajectory 
is an \emph{$(L_F,\kappa)$-loop} if $\w_{0} = (w_0,w_1,\dots,w_{\kappa-1})(w_{\kappa},\dots, w_{L_F})^\omega$, where $\omega$ denotes infinite repetition, and with prefix path length $0 < \kappa \leq L_F$, and equality $w_{\kappa-1} = w_{L_F}$. 
\end{definition} 

If the input-output trajectory $\w_{[0,L_F]} \in \Bfint{[0,L_F]}$ is such that $w_{\kappa-1} = w_{L_F}$, $w_{\kappa} = w_{L_F+1}, \dots, w_{\kappa+\eta-1} = w_{L_F+\eta}$ hold  for some $0 < \kappa \leq L_F$, then we can question under which condition it holds that we get  \emph{$(L_F,\kappa)$-loop}  $\w_{0} = (w_0,w_1,\dots,w_{\kappa-1})(w_{\kappa},\dots, w_{L_F})^\omega \in \Beh$. %, meaning that it belongs to the behavior of the system. 

This is especially of interest since we consider 
a data-driven characterization of the system (see Section~\ref{sec:dataDriven}).  It can easily be shown that it is sufficient to require that the number of input-output pairs that is the same to be at least larger than the lag of the system $\mathbf{l}(\mathfrak{B})$, that is,  \begin{equation}\label{eq:loopConstr}
\eta+1 \geq \mathbf{l}(\mathfrak{B}).
\end{equation} 
In Fig. \ref{fig:loopfigs} the loop constraints on both a state trajectory and input-output trajectory that are $(L_F,\kappa)$-loops are shown.
\color{black}

Next, we construct a bounded specification with additional constraints, such that a satisfying (finite) trajectory implies that the unbounded specification is also satisfied. Denote such a bounded specification by $\varphi_{F}$ and its horizon by $L_F+1$. 
With some abuse of notation, we say that $\w_0= (\u_0,\y_0)$ satisfies a specification, i.e. $\w_0 \models \varphi$ iff for its output trajectory we have $\y _0\models \varphi$. 
We can now conclude the following. 
\begin{thm} \label{th:infHor2}
		Given dynamical system $\Sys=(\Z,\W,\Beh)$, unbounded STL specification $\varphi$, bounded specification $\varphi_F$, and $L_F$.  
		If the input-output trajectory $\w_{[0,L_F]} \in \Bfint{[0,L_F]}$ is such that $w_{\kappa-1} = w_{L_F}$, $w_{\kappa} = w_{L_F+1}, \dots, w_{\kappa+\eta-1} = w_{L_F+\eta}$ %hold  for some $0 < \kappa \leq L_F$ and 
		with $\eta$ satisfying \eqref{eq:loopConstr}, then %$\exists L_F, $ such that 
		we get the following implication on the prefix $\w_{[0,L_F]}$ of $\w_0$:
\begin{equation}\label{eq:impl1}
	 \forall \w_{[0,L_F]} \models \varphi_F \implies \w_0 \models \varphi,
\end{equation} where $\w_0$ is an $(L_F,\kappa)$-loop according to Def.~\ref{def:Lkloop}.
\end{thm}
\begin{pf} 
		 \ifthenelse{\equal{\showlongversion}{1}}{
In line with Def.~\ref{def:Lkloop}, the unbounded trajectory $\w_0$ is constructed based on $\w_{[0,L_F]}$ by concatenating the prefix path $(w_0,w_1,\dots,w_{\kappa-1})$ with the loop  $(w_\kappa, \dots, w_{L_F})$ that is repeated infinitely often. Since for $\w_{[0,L_F]}$ we have that \eqref{eq:loopConstr} holds, by construction, this also holds for $\w_0$. Condition  \eqref{eq:loopConstr} implies that we can reconstruct an equivalent state trajectory of the system. Therefore, the equalities $w_{\kappa-1} = w_{L_F}$, $w_{\kappa} = w_{L_F+1}, \dots, w_{\kappa+\eta-1} = w_{L_F+\eta}$ make sure that the state equality as in 
 \cite{wolff2014optimization} is satisfied and an $(L_F, \kappa)$-loop can be constructed. The proof of implication~\eqref{eq:impl1} follows from the reasoning in \cite{wolff2014optimization}.
}
{The proof is given in the online version \cite{}.
}
\end{pf}
}

The optimization problem \eqref{eq:Opt} can now be written as
\begin{subequations}\label{eq:optProb}
\begin{align}
\argmin\limits_{\u_{[0,L_F+\eta]} \in \mathbb{U}^{L_F+\eta+1}, \boldsymbol{\zeta}_t^{\varphi_{F}}} \hspace{-1cm}&\hspace{1cm}J(\u_{[0,{L_F}+\eta]},\y_{[0,{L_F}+\eta]}) \\
\text{s.t.} &  	\begin{bsmallmatrix}\Hnkl_{T_\text{ini}+{L_F}+\eta+1}(\u^\text{data}) \\ \Hnkl_{T_\text{ini}+{L_F}+\eta+1}(\y^\text{data}) \end{bsmallmatrix} \alpha = \begin{bsmallmatrix} \u^\text{ini} \label{eq:constr1}\\ \u_{[0,{L_F}+\eta]} \\ \y^\text{ini} \\ \y_{[0,{L_F}+\eta]} \end{bsmallmatrix} \vspace{2pt} \\
% & \Birgit{y_{{L_F}-\mathbf{l}(\mathfrak{B})} = y_{L_F}} \\
& \boldsymbol{\zeta}_t^{\varphi_F} = 1 \text{ as in \cite{raman2014model}}, \label{eq:constr2} \\
&\zeta_0^{\varphi_F} = 1, \label{eq:optProb:constr3} \\ 
& w_{\kappa-1} = w_{{L_F}}, w_{\kappa} = w_{{L_F}+1}, \dots \label{eq:loopOpt} \\
& \hspace{1cm}	\dots, w_{\kappa+\eta-1} = w_{{L_F}+\eta}. \notag 
\end{align}
\end{subequations} \
The complete direct data-driven  control synthesis procedure for unbounded STL specifications is summarized in Algorithm~\ref{alg:control}. \Birgit{Note that, in Line~2, we assume we get $||\varphi|| = \infty$, otherwise we can use Algorithm~\ref{alg:controlFin} instead. In the last line, 
we use $\u_{[0,{L_F}+\eta]} =  (u_0,u_1,\dots,u_{\kappa-1})(u_{\kappa},\dots, u_{L_F})(u_{L_F+1}, \dots, u_{L_F+\eta})$ to construct $\u_0 =  (u_0,u_1,\dots,u_{\kappa-1})(u_{\kappa},\dots, u_{L_F})^\omega$, where $\omega$ denotes infinite repetition. }
\newline 

\Birgit{
\begin{algorithm}
\caption{Procedure for unbounded STL specifications
}
\begin{algorithmic}[1]
\State \textbf{Input:} $\w^\text{data}, \w^\text{ini}, \varphi, J$,
\State Compute 
$L$, such that $L+1 > \|\varphi\|$ 
\State Compute (or over-approximate) $\mathbf{l}(\mathfrak{B})$ and choose $\eta$, such that \eqref{eq:loopConstr} is satisfied
\State Find $L_F$, $\varphi_F$ and $\kappa$, such that \eqref{eq:impl1} holds
\State  Compute $\Hnkl_{T_\text{ini}+L+\eta+1}(\w^\text{data})$ for constraint \eqref{eq:constr1}
\State Rewrite constraint $\y_{0} \models \varphi$ to MILP constraints \eqref{eq:constr2}
\State $\u_{[0,L_F+\eta]} \leftarrow $ solve optimization problem \eqref{eq:optProb}
\State Construct $\u_0$ based on $\u_{[0,L_F+\eta]}$
\end{algorithmic}\label{alg:control}
\end{algorithm}}

\begin{remark}
\Birgit{Similar to bounded specifications, we can prove that this direct data-driven control synthesis approach is sound (see Thm.~\ref{th:sound}). 
}	
\end{remark}

\section{Case studies}\label{sec:resultss}
In this section, we apply our approach to two case studies and evaluate the results. \Birgit{We first consider a case study with an unbounded specification, followed by one with a bounded specification.} All simulations have been performed on a computer with a 2.3 GHz Quad-Core Intel Core i5 processor and 16 GB 2133 MHz memory. For each case study, we compute the average computation time after performing $10$ simulations and mention the observed standard deviation. The computation time includes all operations, so also includes acquiring data from the data-generating systems. Besides that, we compute the memory usage by considering all data stored in the \texttt{MATLAB} workspace.

\subsubsection{Car platoon}
Inspired by \cite{haesaert2020robust}, we consider a car platooning example with two cars, a leader and a follower.  To design a data-driven controller that controls the distance between the cars, we get data from the data-generating system %with the corresponding matrices 
defined in the Appendix{ of \cite{}}.
Here, the output $y_t$ of the data-generating system equals the distance between the cars. %Furthermore, its three-dimensional state consists of $x_{t,2}$ and $x_{t,3}$ denoting, respectively, the velocity of the follower car and the leader car.  
%We are interested in the distance between the two cars. 
We consider a bounded input $u_t \in \U = [-2,2]$ that influences the velocity of the follower car and therefore, also the distance between the cars. Note that the analytic form of the data-generating system is only used to generate a data  sequence $\w^\text{data}$ of length $31$. We evaluate the results for two different cost functions, $\smash{J_1(\u_0,\y_0) = \|\y_{[0,L]}\|}$ minimizes the distance between the cars, while $\smash{J_2(\u_0,\y_0) = \|\u_{[0,L]}\|}$ minimizes the actuation of the car. 

\begin{figure}[t]
{%
\subfloat[Input $u_t$ at time steps $t$.]{\label{fig:CarResult_u1}%
\includegraphics[width=0.5\columnwidth]{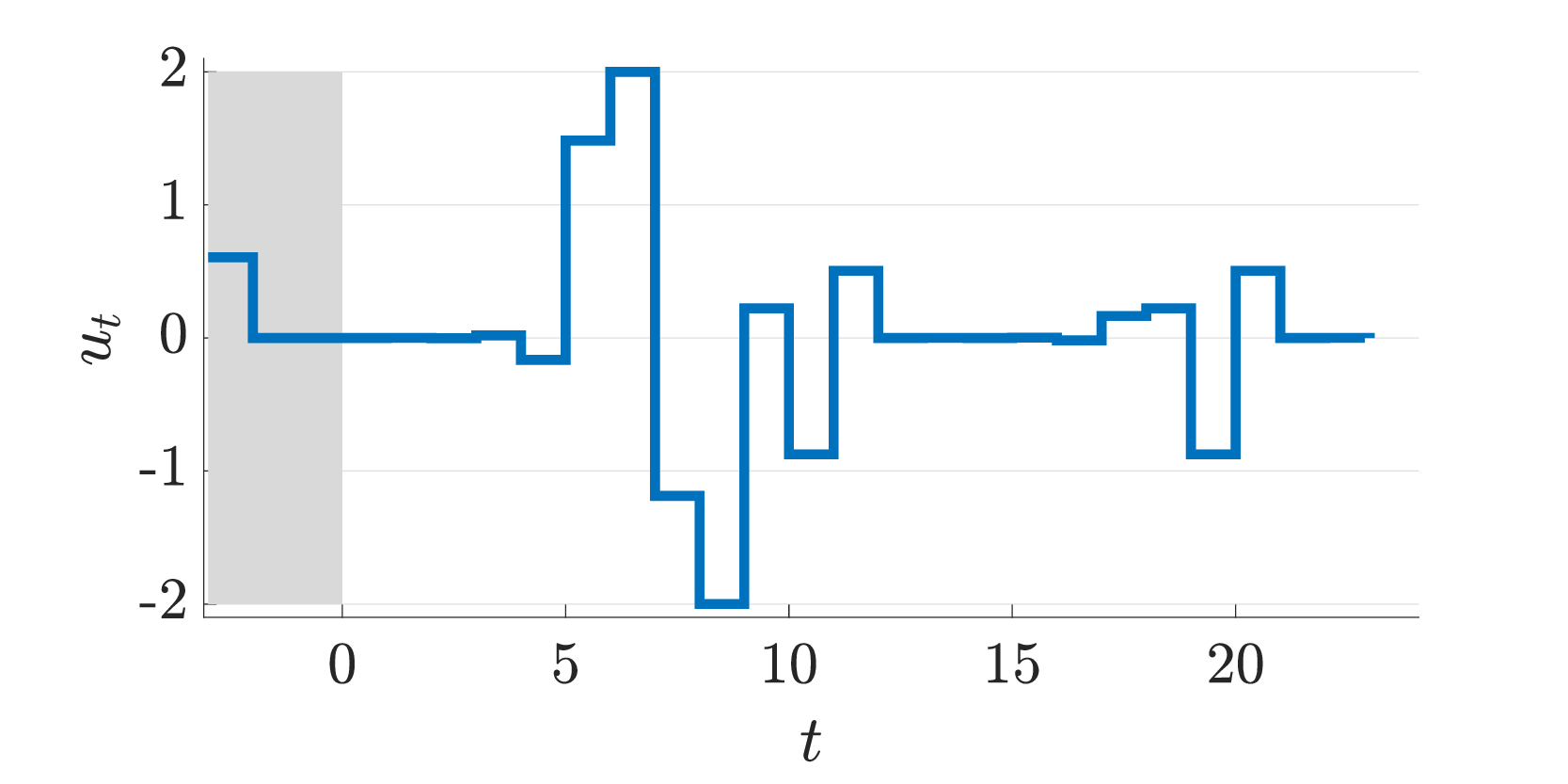}}%
\hfill
\subfloat[Output $y_t$ at time steps $t$.]{\label{fig:CarResult_y1}%
\includegraphics[width=0.5\columnwidth]{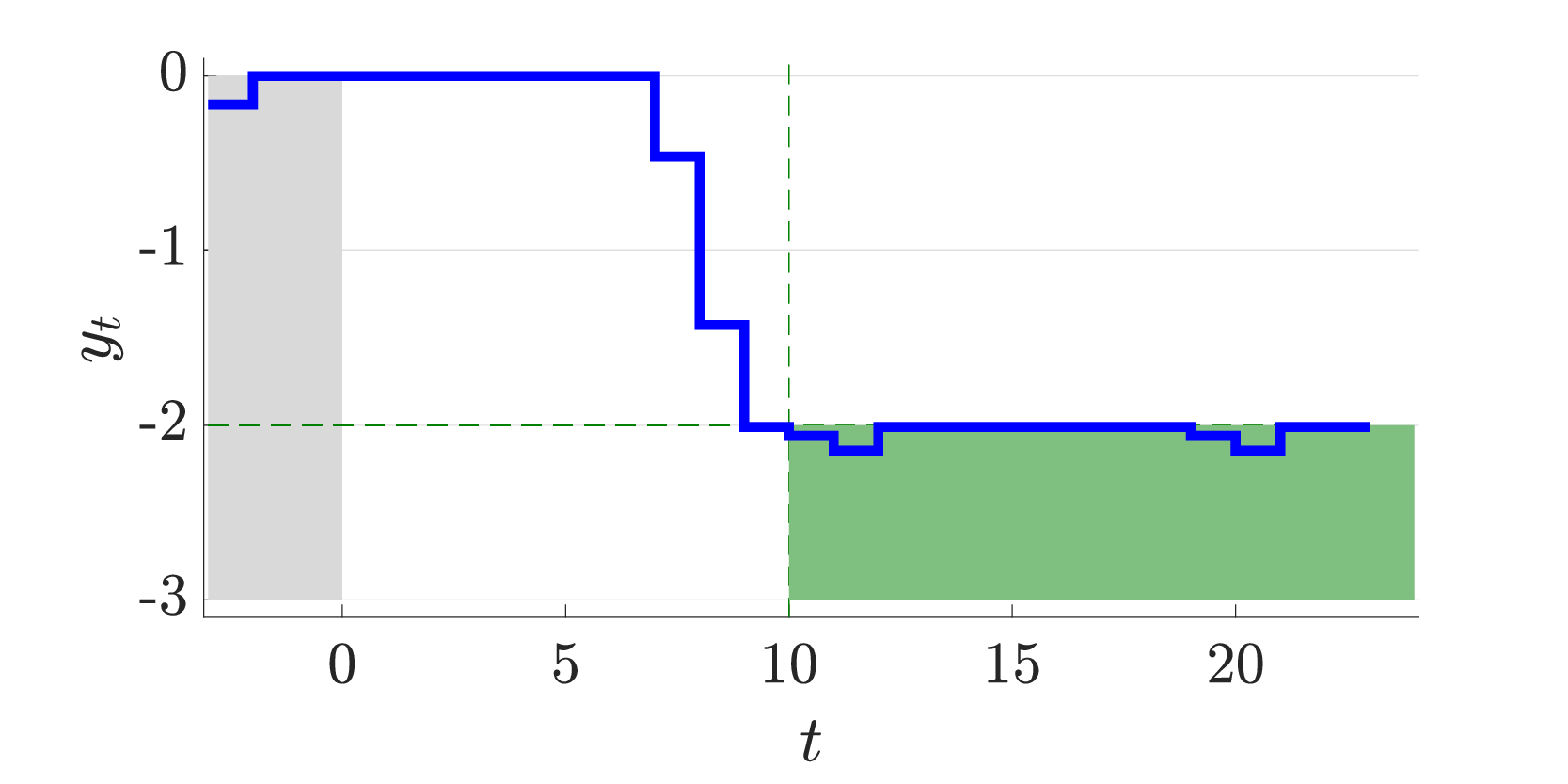}}

\subfloat[Input $u_t$ at time steps $t$.]{\label{fig:CarResult_u2}%
\includegraphics[width=0.5\columnwidth]{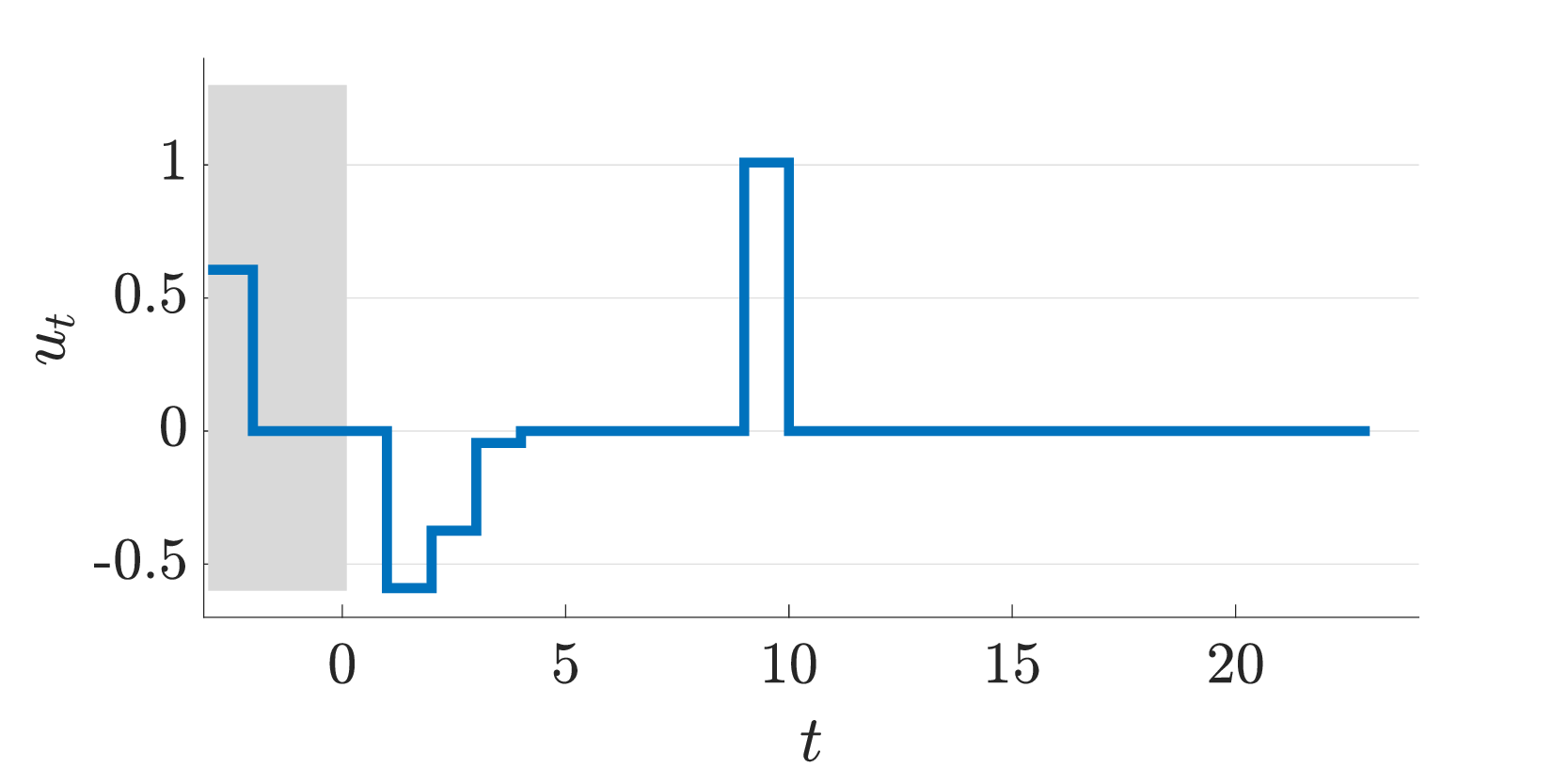}}%
\hfill
\subfloat[Output $y_t$ at time steps $t$ in blue.]{\label{fig:CarResult_y2}%
\includegraphics[width=0.5\columnwidth]{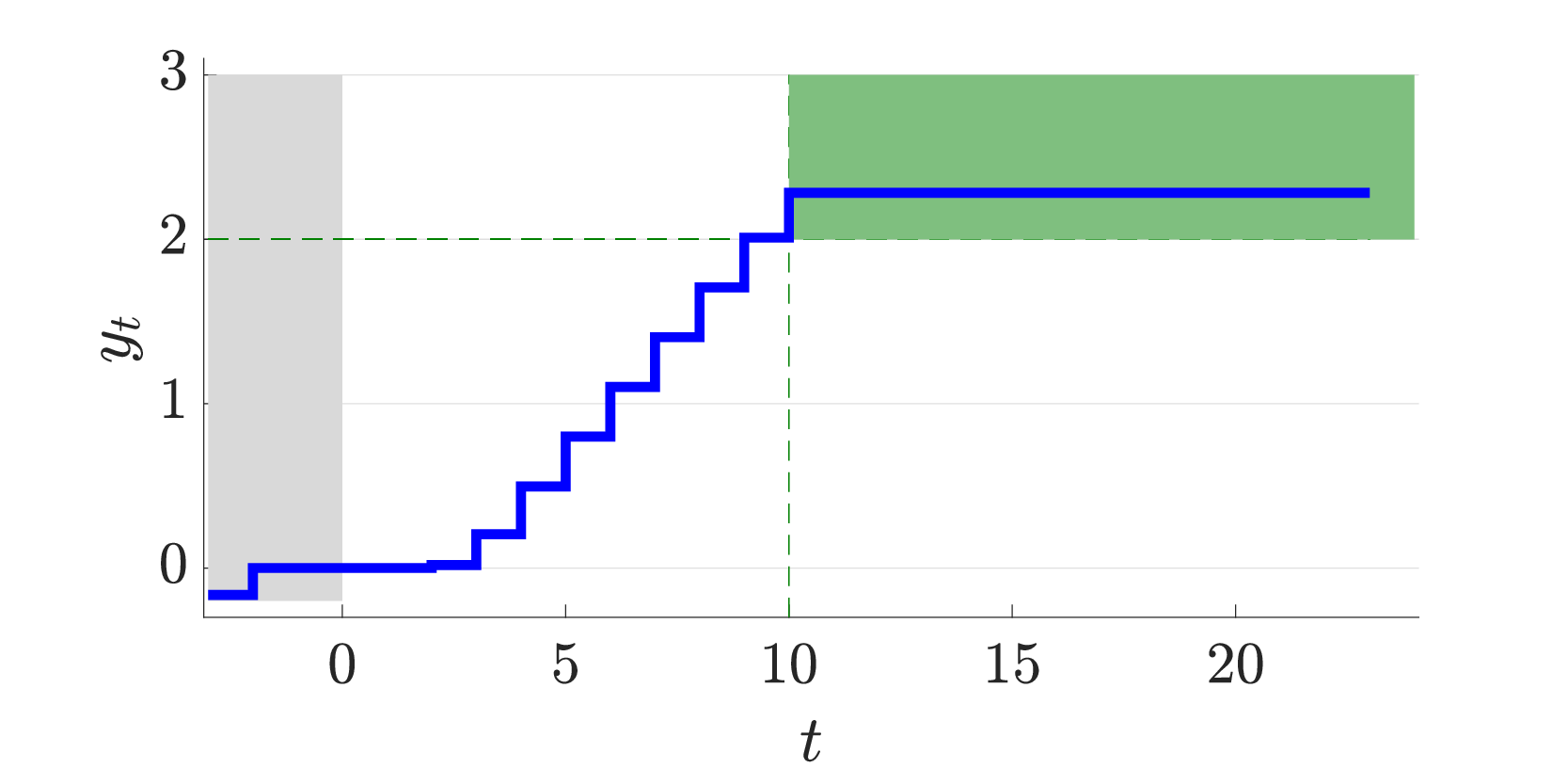}}
}
{\caption{Results of the car platoon problem for cost function $\smash{J_1(\u_0,\y_0)=\|\y_{[0,L]}\|}$ and  $\smash{J_2(\u_0,\y_0)=\|\u_{[0,L]}\|}$, displayed in panels (a), (b) and (c), (d).} 	\label{fig:CarResult}}
\end{figure}

In our considered case, both cars start close to each other with the same constant velocity. The goal of the  controller is to make sure that the follower car achieves a safe distance to the leader car \Birgit{within a specific time frame} and maintains this \Birgit{safe distance forever}.  
This can be written using STL as  
\Birgit{$\varphi = \eventually_{[0,10]} \always (|y| \geq 2 \wedge |y| \leq 3)$.} Note that we did not specify which car should drive in front. After obtaining data $\w^\text{data}$, we have considered initial trajectory  $\w^\text{ini} =(\u^\text{ini} , \y^\text{ini})$ with $\u^\text{ini} = \{0.6058, 0, 0 \}$ and $\y^\text{ini}=\{-0.1636, 0, 0 \}$.
Using Algorithm~\ref{alg:control}, we have obtained the results in Fig.~\ref{fig:CarResult} with $\w^\text{ini}$ highlighted in gray. \Birgit{To construct the looping trajectory, we have chosen $\eta = 4$, $\varphi_F = \eventually_{[0,10]} \always_{[10,19]} (|y| \geq 2 \wedge |y| \leq 3)$, $L_F=19$, and $\kappa = 12$. Note that for these values of $\eta, L_F, \kappa$, the combination of $\varphi_F$ and \eqref{eq:loopOpt} indeed imply the satisfaction of $\varphi$, which is in line with Thm.~\ref{th:infHor2}. For the whole simulation, we observed an average computation time of $7.25$ seconds, with a standard deviation of $0.73$ seconds and a memory usage of $19.78$ MB.}

\Birgit{
In Figs.~\ref{fig:CarResult_y1},~\ref{fig:CarResult_y2}, we clearly see that the specification (indicated by the green box) is satisfied, since the absolute value of the output after time step $\smash{t = 10}$ is between $2$ and $3$. Furthermore, we see the looping trajectory, by comparing the inputs and outputs at time steps $11$ (\smash{$=\kappa-1$}) until $15$ (\smash{$=\kappa+\eta-1$}) with those at $19$ (\smash{$=L_F$}) until $23$ ($\smash{=L_F+\eta}$). Besides that, we conclude that the cost function has a significant influence on the results. In Fig.~\ref{fig:CarResult_y1}, we see that for cost function $J_1$, the controller indeed minimizes the distance between the cars, keeping it as close to $-2$ as possible. Comparing Figs.~\ref{fig:CarResult_u1} and \ref{fig:CarResult_u2}, we can conclude that for cost function $J_2$ the input is substantially smaller over the complete horizon than when cost function $J_1$ is considered.  
}

\subsubsection{Temperature control in a building}
As a second case study, we have considered controlling the temperature inside a building based on the model in \cite{haghighi2013controlling}, where the continuous-time thermodynamics of the multi-room building are approximated with an analogous model of an electric circuit.
Details about the model can be found in \cite{haghighi2013controlling,raman2014model}. After performing a time-discretization (zero-order hold) with a sampling time of $30$ minutes, we have obtained the data-generating system %with the corresponding matrices 
defined in the Appendix\ifthenelse{\equal{\showlongversion}{1}}{}{ of \cite{}}.
The model as in \cite{haghighi2013controlling,raman2014model} has an additional disturbance term $B_d d_t$, which 
we included as a known time-varying term. Note that the analytic form of the data-generating system is only used to generate a data  sequence $\w^\text{data}$ with a length of 1050.

The goal of the controller is to keep the temperature of a room, given by $y_t = T_{t,\text{room}}$ above a certain time-varying comfort level $T_{t,\text{comf}}$ whenever the room is occupied $\text{occ} = 1$, while minimizing the input. We have considered a time horizon of $L+1=48$ (corresponding to $24$ hours). The STL specification is given as $\smash{\varphi_{\text{temp}} = \always_{[0,L+1]} \text{occ} \implies y > T_{t,\text{comf}}}xiv
$ and the cost function is equal to $J(\u_0,\y_0) = \|\u_0\|$. 

We have obtained data $\w^\text{data}$, and initialized with $\w^\text{ini} = (\u^\text{ini},\y^\text{ini})$ with $\u^\text{ini}$ and $\y^\text{ini}$ equal to respectively  $43.65$ and $20$ for $5$ time steps. After converting from degrees Fahrenheit to degrees Celsius, 
we have obtained the results as shown in Fig.~\ref{fig:hvac}. Here, the blue line is the observed output  $y_t = T_{t,\text{room}}$ and the red line is the threshold value for the temperature. This is obtained by combining the occupancy with the comfort level, that is $T_{t,\text{ref}} = \text{occ}_t*(T_{t,\text{comf}})$, with $\text{occ}_t=1$ if the room is occupied and $0$ otherwise. In Fig.~\ref{fig:hvac}, we can see that the specification $\varphi_{\text{temp}}$ is satisfied, since $T_{t,\text{room}} \geq T_{t,\text{ref}}$. For this case study, we observed an average computation time of $26.99$ seconds, with a standard deviation of $7.9$ seconds and a memory usage of $81.1$ MB.

\begin{comment}
\begin{figure}
\centering
\includegraphics[width=.7\columnwidth]{Figures/hvac_temp2}
\caption{Room temperature % (in degrees Celsius) 
	at a specific time. % for the temperature control case study. 
	The output $T_{room}$ is given in blue and the comfort level combined with the occupancy is given in red. }
\label{fig:hvac}
\end{figure}
\end{comment}

\begin{figure}
	\begin{minipage}[c]{0.6\columnwidth}
		\includegraphics[width=\columnwidth]{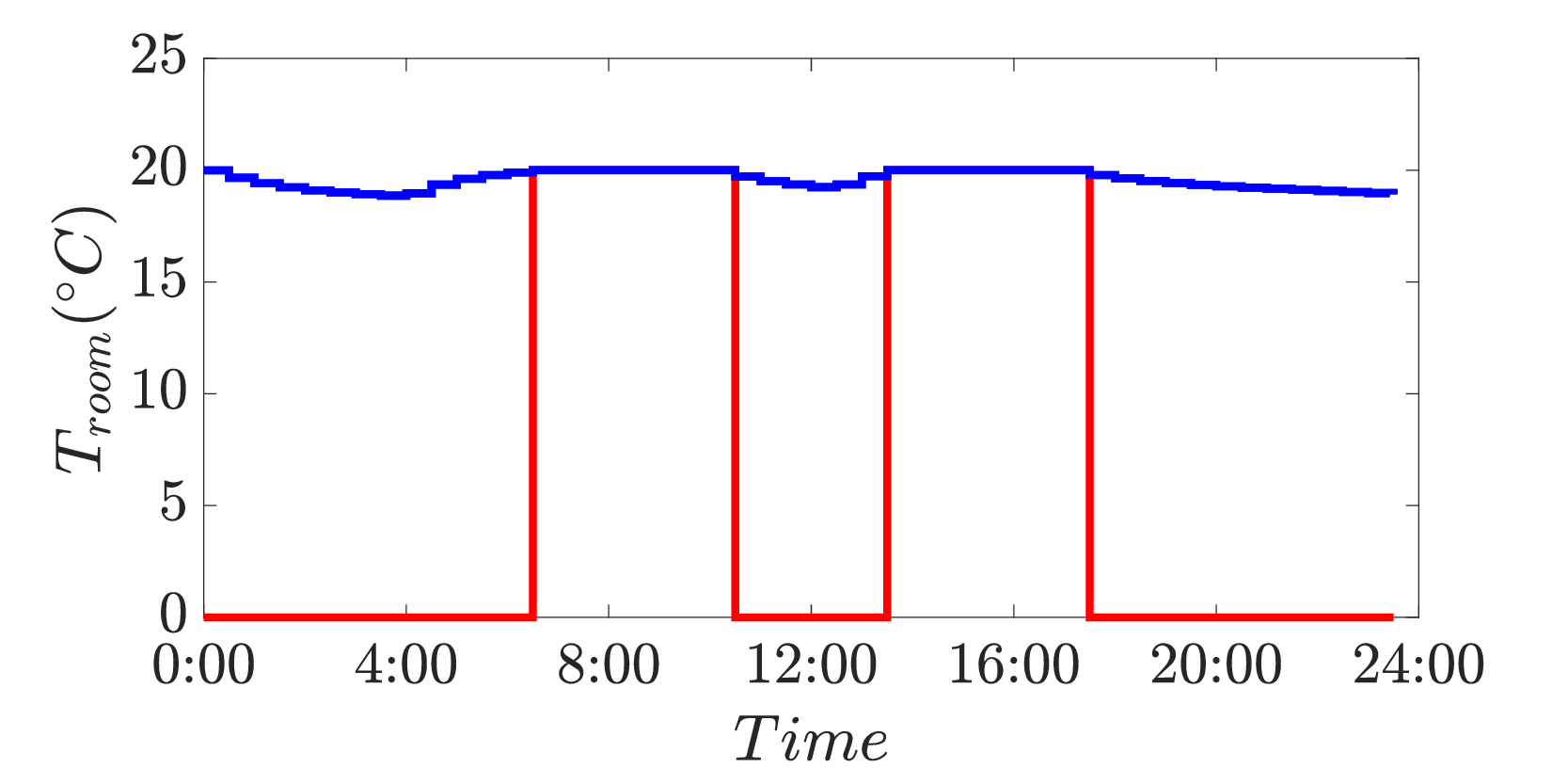}
	\end{minipage}\hfill
	\begin{minipage}[c]{0.4\columnwidth}
		\caption{
		Room temperature (blue) % (in degrees Celsius) 
		at a specific time. % for the temperature control case study. 
		%Output $T_{room}$ is given in blue and t
		The comfort level combined with the occupancy is given in red. } \label{fig:hvac}
	\end{minipage}
\end{figure}

\section{Conclusion}\label{sec:discussion}
In this paper, we have developed a direct data-driven controller synthesis method for linear time-invariant systems subject to a temporal logic specification, which does not require an explicit modeling step. We build upon the promising results of the behavioral framework and the Fundamental Lemma, which allows us to obtain a characterization of the system after collecting a single sequence of input-output data from it. By exploiting existing results on rewriting signal temporal logic specifications to MILP constraints, we can efficiently solve an optimization problem to automatically synthesize a control policy that ensures 
that the specifications are satisfied during closed-loop operation of the system. \Birgit{By constructing a finite trajectory followed by a loop we can even handle unbounded specifications.}
We have 
applied the algorithms to multiple benchmark simulation examples to show that they work well for bounded and unbounded STL specifications. The main benefit of this method is that it only requires input-output data of the system and no exact knowledge of the structure of the system itself. 

\bibliography{Sources}

	 \ifthenelse{\equal{\showlongversion}{1}}{
\appendix
\section{Models used to generate data}
To generate the data required for the case studies, we used the following equations for the car platoon dynamics.
\begin{equation}\label{eq:carMod}
	\begin{array}{l l}
		x_{t+1} &= A_1x_t +B_1u_t \\ 
		y_t &= C_1 x_t,
	\end{array}
\end{equation} 
with 
\begin{equation} \notag \begin{array}{llll}
		A_1 = \begin{bmatrix}
			1 & -0.3 & 0.3 \\
			0 & 1 & 0 \\
			0 & 0 & 1
		\end{bmatrix}
		
		B_1  =  \begin{bmatrix}
			-0.03 \\ 1 \\ 0
		\end{bmatrix}
		
		C_1  = \begin{bmatrix}
			1 \\ 0 \\ 0
		\end{bmatrix}^\top.
	\end{array}
\end{equation}

We used the following equations for the temperature control problem. 
\begin{equation}\label{eq:modelCase}
	\begin{array}{lll}
		x_{t+1} & = & A_2x_t+ \begin{bmatrix} B_2 & B_d \end{bmatrix} \begin{bmatrix} u_t\\  d_t \end{bmatrix}\\
		y_t & =& C_2x_t,
	\end{array}
\end{equation} with 
\begin{equation} \notag \begin{array}{llll}
		A_2 &=\begin{bsmallmatrix}
			0.9233 & 0.00135 & 0.0009377 & 0.002662 & 0.03775 \\
			0.0009377  & 0.9606 &  0.0004754 &  0.00135 & 0.01928 \\
			0.0009377  & 0.0006846 &   0.9604  & 0.00135 & 0.01928 \\
			0.001849  &  0.00135 &  0.0009377 & 0.9241 & 0.03775\\
			0.07636  &  0.05617   &   0.039  &    0.11 & 0.7142 \\
		\end{bsmallmatrix} \vspace{4pt} \\
		
		B_2 & =  \begin{bsmallmatrix}
			3.1194\cdot 10^{-4} \\
			1.5815\cdot 10^{-4} \\
			1.5815 \cdot 10^{-4}\\
			3.1194 \cdot 10^{-4}\\
			0.0131 
		\end{bsmallmatrix}, 
		
		C_2  =  \begin{bmatrix}
			0 & 0 & 0 & 0 & 1
		\end{bmatrix},
	\end{array}
\end{equation} and
\begin{equation}\notag
	\begin{array}{l}
		B_d  = \\
		\begin{bsmallmatrix}
			-8.0390\cdot 10^{-6}  &  0.0340 &   1.9696\cdot 10^{-5} &   3.2720\cdot 10^{-5} &   0.0014  &       0  &       0 \\
			-4.0756\cdot 10^{-6}   & 1.1479\cdot 10^{-5}  &  0.0173   & 1.6530\cdot 10^{-5}   & 0.0230     &    0    &     0 \\
			-4.0756\cdot 10^{-6}   & 1.1479\cdot 10^{-5}    & 0.0173    &  1.6530\cdot 10^{-5}     & 0.0007      &   0     &    0 \\
			-8.0390\cdot 10^{-6}   & 2.2722\cdot 10^{-5}   & 1.9696\cdot 10^{-5}        &0.0340   & 0.0014       &  0      &   0 \\ 
			-3.3691\cdot 10^{-4}   & 0.0014   & 0.0011    &0.0021   & 0.0568        & 0        & 0
		\end{bsmallmatrix} 
	\end{array}
\end{equation}
}{}

\end{document}

%% file: Commands.tex
\newcommand{\N}{\mathbb{N}}

\newcommand{\always}{\Box}
\newcommand{\eventually}{\Diamond}
\newcommand{\until}{\mathbin{\sf U}}

\newcommand{\Hnkl}{\mathcal{H}}
\newcommand{\Beh}{\mathfrak{B}}
\newcommand{\Bfint}[1]{\left.\Beh \right|_{#1}}
\newcommand{\Sys}{\boldsymbol\Sigma}
\newcommand{\T}{\mathbb{T}}
\newcommand{\W}{\mathbb{W}}
\newcommand{\Z}{\mathbb{Z}}

\newcommand{\q}{\mr{q}}

\newcommand{\mf}[1]{\mathfrak{#1}}
\newcommand{\mr}[1]{\mathrm{#1}}
\newcommand{\mb}[1]{\mathbb{#1}}

\newcommand{\mbf}[1]{\mathbf{#1}}

\newcommand{\dnx}{n_\mr{x}}
\newcommand{\dnu}{n_\mr{u}}
\newcommand{\dny}{n_\mr{y}}

\newcommand{\U}{\mathbb{U}}
\newcommand{\Y}{\mathbb{Y}}
\newcommand{\y}{\mathbf{y}}
\renewcommand{\u}{\mathbf{u}}

\newcommand{\w}{\mathbf{w}}

%% file: loopFigure.tex
% !TeX root = l4dc-2024-ddst

\begin{tikzpicture}[->,>=stealth',shorten >=1pt,auto,node distance=2cm,semithick]
	
	% Nodes in a line
	\foreach \i/\label in {1/$x_1$,2/$x_2$,3/$\ldots$,4/$x_{\kappa-1}$}
	\node (L\i) at ( \i-5.5 ,-.5) {\label};
	
	% Nodes in a smaller cycle
	\foreach \i/\label in {1/   ,2/$x_{\kappa+1}$,3/$\ldots$,4/$\ldots$, 5/$x_{L-2}$, 6/$x_{L-1}$}
	\node[yshift=.6cm] (C\i) at (220 + \i * 60:1.4 and 1) {\label};

	% Draw edges between nodes in a line
	\foreach \i in {1,...,3}
	\draw (L\i) -- (L\the\numexpr\i+1\relax);
	
	% Draw edges between nodes in a smaller cycle
	\foreach \i in {1,...,5}
	\draw (C\i) -- (C\the\numexpr\i+1\relax);
			\node[rectangle, fill=black!10,minimum height=.8cm] (C0) at (C1) {$x_\kappa=x_L$};
	% Draw edge from last node in line to the first node in the smaller cycle
	\draw (L4) -- (C0);
	
	% Connect the first and last nodes in the smaller cycle
	\draw (C6) -- (C0.west);
	\draw (C0.east) -- (C2.south);
\end{tikzpicture}

%% file: loopFigure_lag.tex
% !TeX root = l4dc-2024-ddst
\begin{tikzpicture}[->,>=stealth',shorten >=1pt,auto,node distance=0cm,semithick]
	
	% Nodes in a line
	\foreach \i/\label/\colorb  in {0/$w_1$/white,    1/$\ldots$/white,     2/$w_{\kappa-1}$ \rotatebox[origin=c]{90}{$=$} \\$w_L$/black!10, 3/$\ldots$\\\rotatebox[origin=c]{90}{$=$}\\$\ldots$ \hspace{1cm}/black!10,4/$w_{\kappa+\eta-1}$ \rotatebox[origin=c]{90}{$=$} $w_{L+\eta}$/black!10}
	\node[  xshift=-.5cm, rectangle, fill=\colorb,text width=1cm,align=center,minimum height=1cm] (L\i) at ( 1.6* \i-5.5 ,-.5) {\label};
	
	% Nodes in a smaller cycle
	\foreach \i/\label/\nega in {1/$w_{ \kappa+\eta}$/0,2/$w_{\kappa+\eta+1}$/0,3/$\ldots$/0,4/$\ldots$/1, 5/$w_{L-2}$/2}
	\node[yshift=0.4cm, xshift=-2cm,align=center] (C\i) at ( -20 + \i * 40: 2.5 and 1) {\label};
	
	% Draw edges between nodes in a line
	\foreach \i in {0,...,3}
	\draw (L\i) -- (L\the\numexpr\i+1\relax);
	
	% Draw edges between nodes in a smaller cycle
	\foreach \i in {1,...,4}
	\draw (C\i) -- (C\the\numexpr\i+1\relax);
	% Draw edge from last node in line to the first node in the smaller cycle

	\draw (L4) -- (C1);
	
	% Connect the first and last nodes in the smaller cycle
	\draw (C5) -- (L2); 
\end{tikzpicture}
 

%% file: adhs-2024-ddstl2.bbl
\begin{thebibliography}{38}
\providecommand{\natexlab}[1]{#1}
\providecommand{\url}[1]{\texttt{#1}}
\providecommand{\urlprefix}{URL }
\expandafter\ifx\csname urlstyle\endcsname\relax
  \providecommand{\doi}[1]{doi:\discretionary{}{}{}#1}\else
  \providecommand{\doi}{doi:\discretionary{}{}{}\begingroup
  \urlstyle{rm}\Url}\fi

\bibitem[{Baier and Katoen(2008)}]{baier2008principles}
Baier, C. and Katoen, J.P. (2008).
\newblock \emph{Principles of model checking}.
\newblock MIT press.

\bibitem[{Belta et~al.(2017)Belta, Yordanov, and Gol}]{belta2017formal}
Belta, C., Yordanov, B., and Gol, E.A. (2017).
\newblock \emph{Formal methods for discrete-time dynamical systems}, volume~15.
\newblock Springer.

\bibitem[{Berberich and Allg{\"o}wer(2020)}]{berberich2020trajectory}
Berberich, J. and Allg{\"o}wer, F. (2020).
\newblock A trajectory-based framework for data-driven system analysis and
  control.
\newblock In \emph{2020 European Control Conference (ECC)}, 1365--1370. IEEE.

\bibitem[{Berberich et~al.(2020)Berberich, K{\"o}hler, M{\"u}ller, and
  Allg{\"o}wer}]{berberich2020data}
Berberich, J., K{\"o}hler, J., M{\"u}ller, M.A., and Allg{\"o}wer, F. (2020).
\newblock Data-driven model predictive control with stability and robustness
  guarantees.
\newblock \emph{IEEE Transactions on Automatic Control}, 66(4), 1702--1717.

\bibitem[{Breschi et~al.(2023)Breschi, Chiuso, and Formentin}]{breschi2023data}
Breschi, V., Chiuso, A., and Formentin, S. (2023).
\newblock Data-driven predictive control in a stochastic setting: A unified
  framework.
\newblock \emph{Automatica}, 152, 110961.

\bibitem[{Coulson et~al.(2019)Coulson, Lygeros, and
  D{\"o}rfler}]{coulson2019data}
Coulson, J., Lygeros, J., and D{\"o}rfler, F. (2019).
\newblock Data-enabled predictive control: In the shallows of the deepc.
\newblock In \emph{2019 18th European Control Conference (ECC)}, 307--312.
  IEEE.

\bibitem[{De~Persis and Tesi(2019)}]{de2019formulas}
De~Persis, C. and Tesi, P. (2019).
\newblock Formulas for data-driven control: Stabilization, optimality, and
  robustness.
\newblock \emph{IEEE Transactions on Automatic Control}, 65(3), 909--924.

\bibitem[{Deshmukh et~al.(2017)Deshmukh, Donz{\'e}, Ghosh, Jin, Juniwal, and
  Seshia}]{deshmukh2017robust}
Deshmukh, J.V., Donz{\'e}, A., Ghosh, S., Jin, X., Juniwal, G., and Seshia,
  S.A. (2017).
\newblock Robust online monitoring of signal temporal logic.
\newblock \emph{Formal Methods in System Design}, 51, 5--30.

\bibitem[{Devonport et~al.(2021)Devonport, Saoud, and
  Arcak}]{devonport2021symbolic}
Devonport, A., Saoud, A., and Arcak, M. (2021).
\newblock Symbolic abstractions from data: A pac learning approach.
\newblock In \emph{2021 60th IEEE Conference on Decision and Control (CDC)},
  599--604. IEEE.

\bibitem[{D{\"o}rfler(2023)}]{dorfler2023data}
D{\"o}rfler, F. (2023).
\newblock Data-driven control: Part two of two: Hot take: Why not go with
  models?
\newblock \emph{IEEE Control Systems Magazine}, 43(6), 27--31.

\bibitem[{Filippidis et~al.(2016)Filippidis, Dathathri, Livingston, Ozay, and
  Murray}]{filippidis2016control}
Filippidis, I., Dathathri, S., Livingston, S.C., Ozay, N., and Murray, R.M.
  (2016).
\newblock Control design for hybrid systems with tulip: The temporal logic
  planning toolbox.
\newblock In \emph{2016 IEEE Conference on Control Applications (CCA)},
  1030--1041. IEEE.

\bibitem[{{Gurobi Optimization, LLC}(2023)}]{gurobi}
{Gurobi Optimization, LLC} (2023).
\newblock {Gurobi Optimizer Reference Manual}.
\newblock \urlprefix\url{https://www.gurobi.com}.

\bibitem[{Haesaert and Soudjani(2020)}]{haesaert2020robust}
Haesaert, S. and Soudjani, S. (2020).
\newblock Robust dynamic programming for temporal logic control of stochastic
  systems.
\newblock \emph{IEEE Transactions on Automatic Control}, 66(6), 2496--2511.

\bibitem[{Haesaert et~al.(2017)Haesaert, Van~den Hof, and
  Abate}]{haesaert2017data}
Haesaert, S., Van~den Hof, P.M., and Abate, A. (2017).
\newblock Data-driven and model-based verification via bayesian identification
  and reachability analysis.
\newblock \emph{Automatica}, 79, 115--126.

\bibitem[{Haghighi(2013)}]{haghighi2013controlling}
Haghighi, M.M. (2013).
\newblock \emph{Controlling energy-efficient buildings in the context of smart
  grid: A cyber physical system approach}.
\newblock Ph.D. thesis, University of California, Berkeley.

\bibitem[{Hjalmarsson(2005)}]{hjalmarsson2005experiment}
Hjalmarsson, H. (2005).
\newblock From experiment design to closed-loop control.
\newblock \emph{Automatica}, 41(3), 393--438.

\bibitem[{Kalagarla et~al.(2021)Kalagarla, Jain, and
  Nuzzo}]{kalagarla2021model}
Kalagarla, K.C., Jain, R., and Nuzzo, P. (2021).
\newblock Model-free reinforcement learning for optimal control of {M}arkov
  decision processes under signal temporal logic specifications.
\newblock In \emph{2021 60th IEEE Conference on Decision and Control (CDC)},
  2252--2257. IEEE.

\bibitem[{Kapoor et~al.(2020)Kapoor, Balakrishnan, and
  Deshmukh}]{kapoor2020model}
Kapoor, P., Balakrishnan, A., and Deshmukh, J.V. (2020).
\newblock Model-based reinforcement learning from signal temporal logic
  specifications.
\newblock \emph{arXiv preprint}.

\bibitem[{Kazemi et~al.(2022)Kazemi, Majumdar, Salamati, Soudjani, and
  Wooding}]{kazemi2022data}
Kazemi, M., Majumdar, R., Salamati, M., Soudjani, S., and Wooding, B. (2022).
\newblock Data-driven abstraction-based control synthesis.
\newblock \emph{arXiv preprint}.

\bibitem[{Kazemi and Soudjani(2020)}]{kazemi2020formal}
Kazemi, M. and Soudjani, S. (2020).
\newblock Formal policy synthesis for continuous-state systems via
  reinforcement learning.
\newblock In \emph{Integrated Formal Methods (IFM): 16th International
  Conference}, 3--21. Springer.

\bibitem[{Koch et~al.(2021)Koch, Berberich, and
  Allg{\"o}wer}]{koch2021provably}
Koch, A., Berberich, J., and Allg{\"o}wer, F. (2021).
\newblock Provably robust verification of dissipativity properties from data.
\newblock \emph{IEEE Transactions on Automatic Control}, 67(8), 4248--4255.

\bibitem[{Lavaei et~al.(2022)Lavaei, Soudjani, Frazzoli, and
  Zamani}]{lavaei2022constructing}
Lavaei, A., Soudjani, S., Frazzoli, E., and Zamani, M. (2022).
\newblock Constructing {MDP} abstractions using data with formal guarantees.
\newblock \emph{IEEE Control Systems Letters}, 7, 460--465.

\bibitem[{Maler and Nickovic(2004)}]{maler2004monitoring}
Maler, O. and Nickovic, D. (2004).
\newblock Monitoring temporal properties of continuous signals.
\newblock In \emph{Formal Modeling and Analysis of Timed Systems (FORMATS) and
  Formal Techniques in Real-Time and Fault-Tolerant Systems (FTRTFT)},
  152--166. Springer.

\bibitem[{Markovsky and Dörfler(2021)}]{MARKOVSKY202142}
Markovsky, I. and Dörfler, F. (2021).
\newblock Behavioral systems theory in data-driven analysis, signal processing,
  and control.
\newblock \emph{Annual Reviews in Control}, 52, 42--64.

\bibitem[{Markovsky and Rapisarda(2007)}]{markovsky2007linear}
Markovsky, I. and Rapisarda, P. (2007).
\newblock On the linear quadratic data-driven control.
\newblock In \emph{2007 European Control Conference (ECC)}, 5313--5318. IEEE.

\bibitem[{Markovsky and Rapisarda(2008)}]{markovsky2008data}
Markovsky, I. and Rapisarda, P. (2008).
\newblock Data-driven simulation and control.
\newblock \emph{International Journal of Control}, 81(12), 1946--1959.

\bibitem[{Polderman and Willems(1997)}]{willems1997introduction}
Polderman, J.W. and Willems, J.C. (1997).
\newblock \emph{Introduction to mathematical systems theory: a behavioral
  approach}, volume~26.
\newblock Springer Science \& Business Media.

\bibitem[{Raman et~al.(2014)Raman, Donz{\'e}, Maasoumy, Murray,
  Sangiovanni-Vincentelli, and Seshia}]{raman2014model}
Raman, V., Donz{\'e}, A., Maasoumy, M., Murray, R.M., Sangiovanni-Vincentelli,
  A., and Seshia, S.A. (2014).
\newblock Model predictive control with signal temporal logic specifications.
\newblock In \emph{Proc. of the 53rd IEEE Conference on Decision and Control},
  81--87. IEEE.

\bibitem[{Recht(2019)}]{recht2019tour}
Recht, B. (2019).
\newblock A tour of reinforcement learning: The view from continuous control.
\newblock \emph{Annual Review of Control, Robotics, and Autonomous Systems}, 2,
  253--279.

\bibitem[{Romer et~al.(2019)Romer, Berberich, K{\"o}hler, and
  Allg{\"o}wer}]{romer2019one}
Romer, A., Berberich, J., K{\"o}hler, J., and Allg{\"o}wer, F. (2019).
\newblock One-shot verification of dissipativity properties from input--output
  data.
\newblock \emph{IEEE Control Systems Letters}, 3(3), 709--714.

\bibitem[{Sutton and Barto(1999)}]{sutton1999reinforcement}
Sutton, R.S. and Barto, A.G. (1999).
\newblock Reinforcement learning: An introduction.
\newblock \emph{Robotica}, 17(2), 229--235.

\bibitem[{Tabuada(2009)}]{tabuada2009verification}
Tabuada, P. (2009).
\newblock \emph{Verification and control of hybrid systems: a symbolic
  approach}.
\newblock Springer Science \& Business Media.

\bibitem[{Van~Waarde et~al.(2022)Van~Waarde, Camlibel, Rapisarda, and
  Trentelman}]{van2022data}
Van~Waarde, H.J., Camlibel, M.K., Rapisarda, P., and Trentelman, H.L. (2022).
\newblock Data-driven dissipativity analysis: application of the matrix
  s-lemma.
\newblock \emph{IEEE Control Systems Magazine}, 42(3), 140--149.

\bibitem[{Verhoek et~al.(2021{\natexlab{a}})Verhoek, Abbas, T{\'o}th, and
  Haesaert}]{verhoek2021data}
Verhoek, C., Abbas, H.S., T{\'o}th, R., and Haesaert, S. (2021{\natexlab{a}}).
\newblock Data-driven predictive control for linear parameter-varying systems.
\newblock \emph{IFAC-PapersOnLine}, 54(8), 101--108.

\bibitem[{Verhoek et~al.(2021{\natexlab{b}})Verhoek, T{\'o}th, Haesaert, and
  Koch}]{verhoek2021fundamental}
Verhoek, C., T{\'o}th, R., Haesaert, S., and Koch, A. (2021{\natexlab{b}}).
\newblock Fundamental lemma for data-driven analysis of linear
  parameter-varying systems.
\newblock In \emph{2021 60th IEEE Conference on Decision and Control (CDC)},
  5040--5046. IEEE.

\bibitem[{Wang et~al.(2020)Wang, Nair, and Althoff}]{wang2020falsification}
Wang, X., Nair, S., and Althoff, M. (2020).
\newblock Falsification-based robust adversarial reinforcement learning.
\newblock In \emph{2020 19th IEEE International Conference on Machine Learning
  and Applications (ICMLA)}, 205--212. IEEE.

\bibitem[{Willems et~al.(2005)Willems, Rapisarda, Markovsky, and
  De~Moor}]{willems2005note}
Willems, J.C., Rapisarda, P., Markovsky, I., and De~Moor, B.L. (2005).
\newblock A note on persistency of excitation.
\newblock \emph{Systems \& Control Letters}, 54(4), 325--329.

\bibitem[{Wolff et~al.(2014)Wolff, Topcu, and Murray}]{wolff2014optimization}
Wolff, E.M., Topcu, U., and Murray, R.M. (2014).
\newblock Optimization-based control of nonlinear systems with linear temporal
  logic specifications.
\newblock In \emph{International Conference on Robotics and Automation},
  5319--5325.

\end{thebibliography}
